\documentclass{IJCAS}

\usepackage{url}
\usepackage{marvosym}
\usepackage{amsmath,amssymb,amsfonts}
\usepackage{array,tabularx}
\usepackage{multicol} 

\usepackage{subfigure}
\usepackage{graphicx}
\usepackage{booktabs} 
\usepackage{enumerate}

\journalvolumn{VV}
\journalnumber{X}
\journalyear{YYYY}
\setarticlestartpagenumber{1}

\allowdisplaybreaks

\begin{document}

\title{AGENT: An Adaptive Grouping Entrapping Method of Flocking Systems}

\author{Chen Wang\orcid{}, Minqiang Gu\orcid{}, Wenxi Kuang\orcid{}, Dongliang Wang\orcid{}, Weicheng Luo\orcid{}, Zhaohui Shi\orcid{}, Zhun Fan*\orcid{}
}

\begin{abstract}
		This study proposes a distributed algorithm that makes agents’ adaptive grouping entrap multiple targets via automatic decision making, smooth flocking, and well-distributed entrapping. Agents make their own decisions about which targets to surround based on environmental information. An improved artificial potential field method is proposed to enable agents to smoothly and naturally change the formation to adapt to the environment. The proposed strategies guarantee that the coordination of swarm agents develops the phenomenon of multiple targets entrapping at the swarm level. We validate the performance of the proposed method using simulation experiments and design indicators for the analysis of these simulation and physical experiments.  
\end{abstract}


\maketitle

\makeAuthorInformation{
Manuscript received January 10, 2022; revised March 10, 2022; accepted May 10, 2022. Recommended by Associate Editor Soon-Shin Lee under the direction of Editor Milton John. This journal was supported by the Korean Federation of Science and Technology Societies Grant.\\

Chen Wang, Minqiang Gu, Wenxi Kuang, Dongliang Wang, Weicheng Luo, Zhaohui Shi, Zhun Fan are with the College of engineering, Shantou University, Shantou City, Guangdong Province, China \{20cwang2,mqgu,20wxkaung, dlwang,19wcluo,19zhshi1, zfan\}@stu.edu.cn). 


* Corresponding author.
}

\runningtitle{2022}{Chen Wang, Minqiang Gu, Wenxi Kuang, Dongliang Wang, Weicheng Luo, Zhaohui Shi, Zhun Fan*}{AGENT: An Adaptive Grouping Entrapping Method of Flocking Systems}{xxx}{xxxx}{x}

\section{Introduction}

There have been increasing research interests in the distributed cooperative control of multi-agent systems 
generating emergent flocking behaviors. These studies have received considerable attention since Reynolds proposed three heuristic rules \cite{c1} including collision avoidance, velocity matching and flock centering for multi-agent.
Based on the three general rules, hundreds of models have emerged to model the synchronized collective motion of animals, humans, or even migrating cells \cite{c1,c2,c3,c4,c5}.
Applications of swarm systems including search and rescue \cite{c6}, area/border coverage \cite{c7}, deployment of sensor networks \cite{c8}, collective transportation and construction \cite{c9}, and convoy/escorting missions \cite{c10,c11}. 

In robotic distributed swarm systems, target entrapping capture is a typical challenging research field\cite{c12}. 
Much work has been done on target entrapping.
Zhou et al.\cite{c13} introduced the escape-encircle strategy exhibited by biological community into the research of multi-uav cooperative combat, and a strategy of multi-UAV cooperative encircle target was designed.
Yao et al.\cite{c14} proposed circumnavigation control algorithms enable multiple robots to rotate around a target which achieving entrapping effect. However, these methods does not consider to change formation shape of swarm robots adaptively in the environment with obstacles. 
In addition, there are many methods such as: behavior-based control methods\cite{c15,c16}, virtual structure methods\cite{c17,c18}, leader-follower control methods\cite{c19,c20}, and biological heuristic methods\cite{c21,c22}. However, these methods have not solved these problems comprehensively. These algorithms may be less applicable when there are multiple dynamic targets in the environment.
To the best of our knowledge, only a handful of efforts have focused on multiple target entrapping.
Kubo et al.\cite{c23} proposed a swarm robot multi-target entrapping algorithm. However, the multiple targets are stationary.
Yasuda et al.\cite{c24} used swarm robots to entrap and transport multiple targets based on evolutionary artificial neural network united states artificial neural networks (EANNs). But there are no obstacles in the environment. 

The challenge in achieving a large-scale swarm of robots that dynamically surround multiple targets is how to make each robot self-organize tasks when the dynamic multiple targets scatter. This requires the robots in the field to surround each target as evenly as possible through their own decisions; it also requires the system to be robust.
The gene regulatory network(GRN) method has achieved good results in previous research on entrapping.
Inspired by the genetic and cellular mechanisms that control biological morphogenesis, Jin et al. \cite{c21} proposed a hierarchical gene regulatory network method that enables agents to generate entrapping formation according to environmental changes. Peng et al. \cite{c22} proposed an improved GRN for entrapping multiple dynamic targets in an environment containing obstacles. On this basis, Fan et al. \cite{c25} used genetic programming to automatically generate the GRN structure according to the scene and realized a better entrapping effect of the robot on the target in a complex obstacle scene. However, GRN cannot effectively deal with the physical constraints of robots. It is well known that ignoring constraints can reduce the robustness of a system.

Vásárhelyi \cite{c26} presented a flocking model for real drones, and the experiments demonstrated that the induced swarm behavior remained stable under realistic conditions for large flock sizes. It completed the task of flocking with an improved artificial potential field method, which gives us a good inspiration for flocking. 
Inspired by\cite{c26}, this study proposes an \underline{a}daptive \underline{g}rouping \underline{ent}rapping method (AGENT) based on an improved artificial potential field method that considers real machine motion constraints for adaptive grouping and target capture of agent swarms. The main contributions of this study are as follows. 
\begin{itemize}	
	
	\item This study proposes a multiple targets entrapping task model framework combining adaptive decision-making mechanism and artificial potential field method. 
	
	\item The improved artificial potential field method proposed in this study makes agents emerge distributed uniform entrapping of targets with strong robustness. 
	
	\item An extensible agent decision mechanism is proposed for agent adaptive target selection.

	\item In this study, several evaluation indexes were established to evaluate the effect of swarm entrapping multiple targets, and simulation experiments were carried out to compare with GRN method. The real experiments were deployed on E-puck2 platform.
	
\end{itemize}

The remainder of this paper is organized as follows.  In Section~\ref{sec2}, we describe the design of the velocity controller. In Section~\ref{sec3}, the adaptive decision-making method is introduced. In section~\ref{sec4}, we provide the comparative experiments (our method and GRN) and propose statistical indicators for the effect of entrapping to statistically analyze the experimental data to confirm the validity and feasibility of our method. Real-world experiments on the E-puck2 robot platform are presented in Section~\ref{sec5}. Finally, Section~\ref{sec6} concludes the paper.

\section{Velocity control mechanism }\label{sec2}

In this section, we introduce the method of entrapping targets by agents. There are some factors we need to consider. When agents perform tasks in a swarm, they should keep their distance from each other\cite{c26}. When they are too close, they should produce a mutually exclusive velocity term. They should produce the same repulsion velocity term when they are close to the targets to avoid collision. In addition, agents also need to avoid obstacles in a timely manner\cite{c26}. The above mechanisms ensure that agents display a collective pattern. To entrap the target, agents must set up specific mechanisms to stay within a certain distance from the target. Finally, in practical engineering applications, the aforementioned motions must consider the mobility of the robot, so the acceleration of speed controller is limited.

\subsection{Close to the target}
When agents get the target information and aim to entrap the target, their goal is to reach a point at a certain distance from the target. Therefore, agents need to have a velocity term that approaches the target point and achieve a smooth decay of velocity as they approach the target. 
$ D(.) $ as an ideal braking curve has a smooth velocity decay function in space, with constant acceleration at high velocity and an exponential approach in space at low velocity \cite{c26}.
In function $ D(.) $, $r$ is the distance between an agent and the expected stopping point. The $p$ gain determines the crossover point between the two phases of deceleration, and $a$ is the preferred acceleration.

\begin{equation}
	D(r, a, p)= \begin{cases}0 & \text { if } r \leq 0 \\ r p & \text { if } 0 <r p \leq a / p \\ \sqrt{2 a r-a^{2} / p^{2}} & \text { otherwise }\end{cases}
\end{equation}

With this smooth decay curve, agents can implement a velocity decay when approaching the target in Eq.2. This is essentially what one does when pressing the brake pedal of a car. First, the brake pedal is pushed at a high velocity; then the velocity is gradually decreased.
\begin{equation}
	v_{\text {it }}=\left[ v_{\!f}  + C^{\text {t}}  \cdot D\left(r_{\text {it }}-R_{\text {entrap }}, a^{\text {t }}, p^{\text {t }}\right)\right] \cdot {\vec{r_{\text {ti }}}}
\end{equation}

In the equations above, we set the initial velocity $v_{\!f}  $ of the agent. $C^{\text {t }} $ is the preferred common travelling velocity coefficient for all agents approaching targets.
$r_{\text {it}}=\left|r_{i}-r_{t}\right| $ is the distance between agents $i$ and the target. $ a^{\text {t }} $ is the maximum acceleration allowed. Higher values assume that agents can brake quicker. Excessively high values result in the inability of agents to react to excessively large velocity differences in time and thus lead to collisions. $ p^{\text {t }} $is the gain of the optimal braking curve used to determine the maximum allowed velocity difference.
Large values approximate the braking curve to the constant acceleration curve. Small values elongate the final part of the braking (at a low velocity) with decreasing acceleration and smoother stops.
$ {\vec{r_{\text {ti }}}} $ is the direction in which the agent points to the target location. 
$R_{\text {entrap}}$ represents the distance of the stopping point in front of the target according to the entrapping task.

\subsection{Repulsion}

Agents generate velocity terms that move away from each other when the distance between agents is under $ r_{\text {arep}} $, the distance at which the local repulsion kicks in. Larger values create sparser flocks with fewer collisions. 

\begin{equation}
	v_{\text {ij}}^{\text {rep }}= \begin{cases}p_{\text {a}}^{\text {rep }}\!\!\left({r_{\text {arep}}}-{r_{\text {ij}}}\right) \cdot {\vec{r_{\text {ij }}}}  &\!\! \text { if }\left(r_{\text {ij }}<r_{\text {arep}}\right) \\ 0 &\!\! \text { otherwise }\end{cases}
\end{equation}

where $r_{\text {ij }}=\left|r_{i}-r_{j}\right| $ is the distance between agents $i$ and $j$,
$ p_{\text {a}}^{\text {rep}} $ is the linear coefficient of the velocity of repulsion between agents, and
$ {\vec{r_{\text {ij }}}} $ represents the direction of the velocity of acting on agent $i$ from agent $j$ to agent $i$.

Furthermore, repulsion is also used between the agents and targets unidirectionally. If the distance between the agent and the target is under $ r_{\text {trep}}$, then the agents will be far away from the target.
\begin{equation}
	v_{\text {itarget}}^{\text {rep}}=\left\{\begin{array}{lc}
		\!\!\!p_{\text {t}}^{\text {rep}}\left({r_{\text {trep}}}-{r_{\text {it}}}\right) \cdot \  {\vec{r_{\text {it }}}}&\!\!\text {if }\left(r_{\text {it}}<r_{\text {trep }}\!\!\right) \\
		\!\!0 &\!\!\!\!\!\!\!\!\!\!\!\!\!\! \text { otherwise }
	\end{array}\right.
\end{equation}

Similarly, $r_{\text {it}}=\left|r_{i}-r_{t}\right| $ is the distance between agents $i$ and the target. $ p_{\text {t}}^{\text {rep}} $ is the linear coefficient of the repulsion velocity between the agent and the targets. $ {\vec{r_{\text {it }}}} $ represents that the repulsion direction acting on agent $i$ is from the target to agent $i$. Each agent needs to calculate the repulsion velocity term for all targets.

To obtain comprehensive repulsion, we take the vectorial sum of the interaction terms of repulsion introduced in Eq.3 and Eq.4:

\begin{equation}
	v_{\text{i}}^{\text {rep}}=\sum_{j \neq i} v_{\text {ij }}^{\text {rep}}+ \sum_{\text {target}} v_{\text {itarget }}^{\text {rep}}
\end{equation}

\subsection{Interaction with walls and obstacles}

In some practical applications, the task of entrapping targets by an agent needs to be carried out within a certain area. In this study, the flocking motion mechanism of the AGENT method considers the boundary constraints\cite{c27,c28} and obstacles of the arena to adapt to certain tasks.
The targets and agents move in a square arena with walls and obstacles. To better avoid collision with the wall, we assumed that there are virtual agents distributed on the boundary of walls and obstacles. The virtual agent is located at the point closest to the agent on the boundary of the wall or obstacle \cite{c29}.

\begin{equation}
	v_{\text {id }}^{\text {wall}}= \begin{cases}0 &\!\!\!\!\!\!\!\!\!\!\!\!\!\!\!\!\!\!\! \text {if }\left(r_{\text {id}}>=r_{\text {wall}}\right) \\ C^{\text {d }}\cdot\left(v_{\text {id }}-D\left(r_{\text {id }}-r_{\text {wall }}, a^{\text {d}}, p^{\text {d}}\right)\right) \!\cdot \!\ {\vec{v_{\text {id }}}} &\!\!\! \text { otherwise }\!\!\end{cases}	
\end{equation}
where  $C^{\text {d }}$ is the velocity coefficient of the distance from the walls, and $r_{\text {wall}}$ is the safe distance from the agents to the wall; 
$r_{\text {id }}=\left|r_{i}-r_{d}\right| $ is the distance between agents $i$ and the closest point on the boundary of the wall or obstacle(virtual agent's position). Here, larger values cause agents to begin braking at larger distances from the wall. $ a^{\text {d}} $ and $ p^{\text {d}} $ is same as $ a^{\text {t}} $ and $ p^{\text {t}} $ in Eq.2 but for avoiding collisions with walls. $\vec{v_{\text {d }}}$ is virtual agent's velocity which is perpendicular to the wall edge pointing inward in the arena(${v_{\text {id }}}=\left|\vec{v_{i}}-\vec{v_{d}}\right| $). $\vec{v_{\text {id }}}$ represents the unit direction vector of the agent's obstacle avoidance direction which is calculated from the vector difference between virtual agent and agent velocities. This method avoids the local minimum value of the potential field to some extent.

Walls and obstacles are similar in agents' obstacle avoidance. Agents can use the same method to avoid obstacles while entrapping the targets; that is, for each agent and obstacle, the velocity component $ v_{\text {id }}^{\text {obs}}$ can be defined similarly to Eq.7. Parameters such as the minimum distance between the expected agent and the wall $r_{\text {wall}} $ can be changed according to the actual needs to be applied to obstacles.

\subsection{Final equation of desired velocity}

The speed controller needs to consider both of these possible effects, so the above velocity influence items need to be superimposed here. The desired velocity calculated by the algorithm is:

\begin{equation}
	\tilde{v}_{i}^{\text {desire }}=v_{\text{i}}^{\text {rep}}+{v}_{\text {it }}+{v}_{\text {id}}^{\text {wall}}+{v}_{\text {id}}^{\text {obs }}
\end{equation}

To make the method closer to the actual application, a velocity limit term  $ v_\text {limit} $ is introduced. If the obtained velocity term is over the limit, its magnitude is reduced without changing the velocity direction.

\begin{equation}
	\tilde{v}_{i}^{\text {desire }}=\frac{\tilde{v}_{i}^{\text {desire }}}{\left|\tilde{v}_{i}^{\text {desire }}\right|} \cdot \min \left\{\left|\tilde{v}_{i}^{\text {desire }}\right|, v_\text {limit} \right\}
\end{equation}

\section{Adaptive decision making}\label{sec3}

During an entrapping mission that encounters multiple targets with equal significance, it is preferred that the agents are grouped evenly to encircle each target. In distributed systems, agents need to make decisions to surround corresponding targets, and the phenomenon of entrapping appears at the swarm level\cite{c30}. In the AGENT method, the problem of target grouping is transformed into the problem of agents selecting targets according to environmental factors.

In the task of entrapping multiple targets, the environmental factors to be considered include the number of agents surrounding the target and the relative distance between the agent and the target.
If the agent is too far from one of the targets relative to the others, or there are already enough agents surrounding the target; consequently, the agent no longer needs to entrap the target. That is to say, the agent usually needs to combine these two factors including the distance from the target and the number of agents surrounding the target in entrapping scenes. Thus, agents should be endowed with the following mechanisms:
The agents calculate their distance from various targets in real-time and detect the number of agents surrounding each target. All agents calculate the $Seq$ matrix, as shown in Eq.9. In this manner, agents make decisions to divide themselves into different groups to entrap different targets autonomously. The correlation factor is considered on the right side of the equation, including the distances from the agent to each target $r_{itn}$ and the number of agents surrounding each target $N_{itn}\ $ in real-time.

\begin{equation}
	(Seq_{1}\ Seq_{2}\,...\ Seq_{n}) = (a\ b)\cdot
	\left (
	\begin{array}{cccc}
		\!\! r_{it1}\!\!&r_{it2}\!\!&...\!\!&r_{itn}\!\!\!\\ 
		\!\! N_{it1}\!\!&N_{it2}\!\!&...\!\!&N_{itn}\!\!\! 
	\end{array}
	\right)
\end{equation}
where $ (a,b) $ is the weight matrix that represents the importance of the two factors in the matrix. We can obtain a matrix representing the target entrapping sequences for each agent. Each agent only needs to entrap the target corresponding to the element sequence with the smallest $Seq$ value in the $Seq$ matrix. Furthermore, the $ Seq $ matrix is updated in time by each agent. Thus, agents make more suitable decisions for efficient entrapping.

We can continue to increase the parameters and corresponding weights to meet the actual needs such that different targets have different importance, $ c $ represents the weight of target's importance. For example, in Eq.10, $ P_{itn} $ represents the different encirclement priorities of different targets. In this manner, the grouping algorithm in the AGENT method can be flexibly applied to a variety of scenes.

\begin{equation}
	\!\!\!\!(Seq_{1}\ Seq_{2}\,...\ Seq_{n}) = (\!a\ b\ ... c\ )\cdot\!
	\left (
	\begin{array}{cccc}
		\!\!\!\!\! r_{it1}\!\!\!&r_{it2}\!\!&\!\!...\!\!\!\!&r_{itn}\!\!\!\!\!\\ 
		\!\!\!\!\! N_{it1}\!\!\!&N_{it2}\!\!&\!\!...\!\!\!\!&N_{itn}\!\!\!\!\!\\
		\!\!\!\!\! ...\!\!\!&...\!\!&\!\!...\!\!\!\!&\!...\!\!\!\!\!\\ 
		\!\!\!\!\! P_{it1}\!\!\!&P_{it2}\!\!&\!\!...\!\!\!\!&P_{itn}\!\!\!\!\!
		
	\end{array}
	\right)
\end{equation}

As the decision-making framework shows, the factors and the weight of each factor that we need to consider in decision-making can be increased according to the actual situation. This parameter matrix can be considered as the weight matrix of neural network. If there are many factors to consider, the method of deep neural network can be considered, which is just like classifying pictures according to pixel values. The dimensions of the weight matrix can be adjusted according to actual conditions. Therefore, the framework exhibits good migration and scalability.

\section{Simulation experiment and analysis}\label{sec4}

\subsection{Simulation experiments}

In this section, the performance of the proposed AGENT method is evaluated using simulation cases based on MATLAB. To demonstrate the validity and robustness of the AGENT method, we set different complex obstacles in a square scene, as shown in Fig.1(a) and Fig.1(b).
The simulation experiment arena (250m*250m) was as follows: There were some agents in blue color and a target in orange color. The mission of the agents was to entrap the target and not crash into other agents, obstacles, or walls. The agents obtained the position information of each other through communication and detected the positions of obstacles and targets. The velocity of the target was $ 2.6 m/step $ and the velocity of the agents was $ 0-4 m/step $. The trajectory of the target and agents is depicted in Fig.1 along with pictures of the key moments when agents entrapped the target.

\begin{figure}[htbp]
	\centering
	
	\subfigure[]{		\label{fig1a}
		\includegraphics[width=0.46\linewidth]{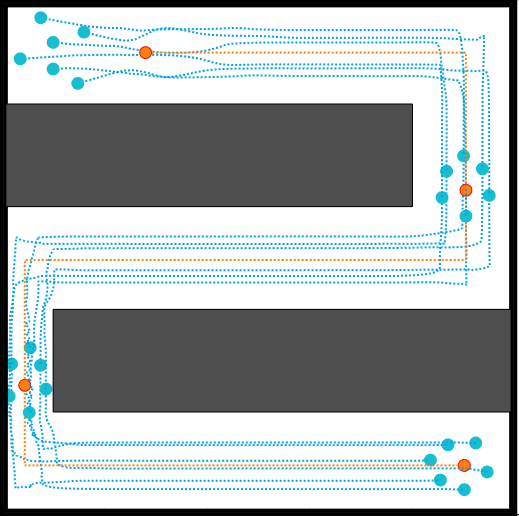}}
	\hspace{0.01\linewidth}
	\subfigure[]{		\label{fig1b}
		\includegraphics[width=0.46\linewidth]{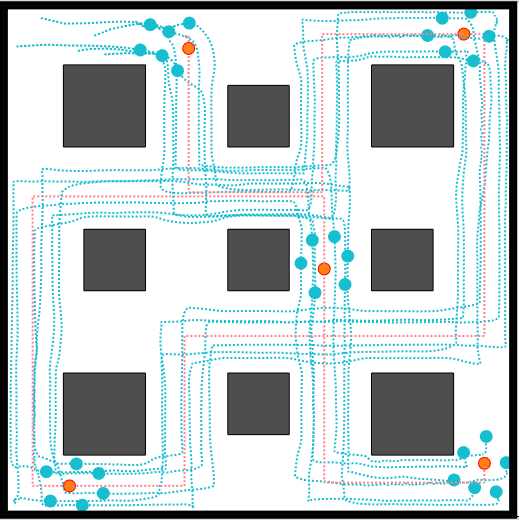}}
	\hspace{0.01\linewidth}
	\caption{Process of agents entrap the target in different scenes. (a) Scene 1. (b) Scene 2. } 	
\end{figure}\label{fig1}

In various complex obstacle scenes, agents can flexibly avoid obstacles and other agents, even in very narrow spaces. It can be seen from the trajectory of the agents that they avoid collision with obstacles in the arena during the entire entrapping process. Regardless of how the target changes direction, agents can entrap the target with good performance. 

To further demonstrate the adaptive decision ability of the AGENT method, we design multiple target entrapping scenes.
Targets wandered in the arena. To make the agents and targets more real, we used the L\'{e}vy flight as the target moving algorithm. The L\'{e}vy distribution is a probability distribution proposed by French mathematician L\'{e}vy in the 1930s. L\'{e}vy flight is a random search path that obeys the L\'{e}vy distribution. This is a random walking mode that alternates between short and long-distance searches. After much research, L\'{e}vy flight conforms to the behavior trajectories of many natural creatures, such as bees and albatross. It can explain many random phenomena in nature. 
%
%
%

The simulation experiment scene was as follows: There were red and green targets in the scene (250 m*250 m). The velocity of the target is $ 2.6 m/step $, and the velocity of the agents was $ 0-4 m/step $. The agent changed its color as it approached the target of the corresponding color similar to a chameleon.

To prove the superiority of our method, the AGENT and GRN methods are compared experimentally in Fig.2 and Fig.3. The GRN method is famous for entrapping with better performance with agents flexibly overcoming obstacles to entrap the targets. The GRN method consists of two layers: the upper layer is for adaptive pattern generation, which is evolved by basic network motifs with genes and environmental inputs, and it can generate a suitable pattern for the dynamic changes of the entrapping target. The lower layer drives the robots to the target pattern generated by the upper layer. The velocity of the agent does not change the amplitude ($1.6 m/step$ was chosen for a better experimental effect in the following experiment) but changes the direction. Owing to the different velocity control mechanisms of the two methods, we compare the distance used for entrapping in the following index analysis for the sake of fairness. The simulation renderings of the agents entrapping two targets in the same scene are as follows (see our simulation experiment video link in the appendix).

\begin{figure*}[htbp]
	\centering
	
	\subfigure[]{
		\includegraphics[width=0.2\linewidth]{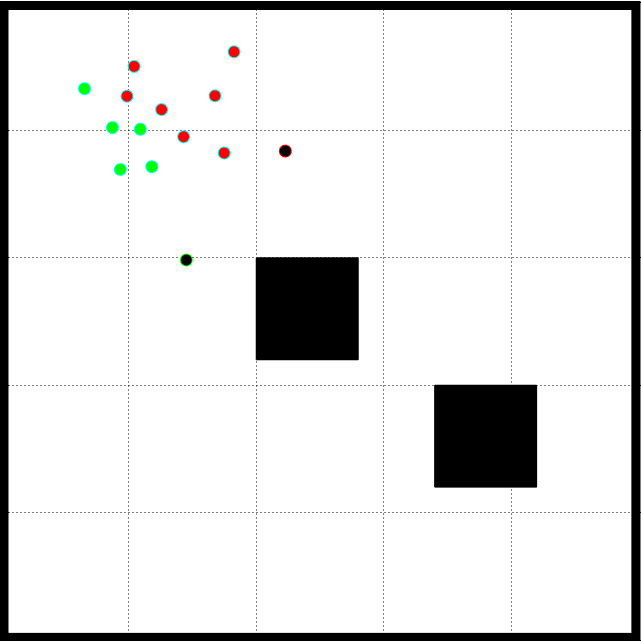}}
	\hspace{0.01\linewidth}
	\subfigure[]{
		\includegraphics[width=0.2\linewidth]{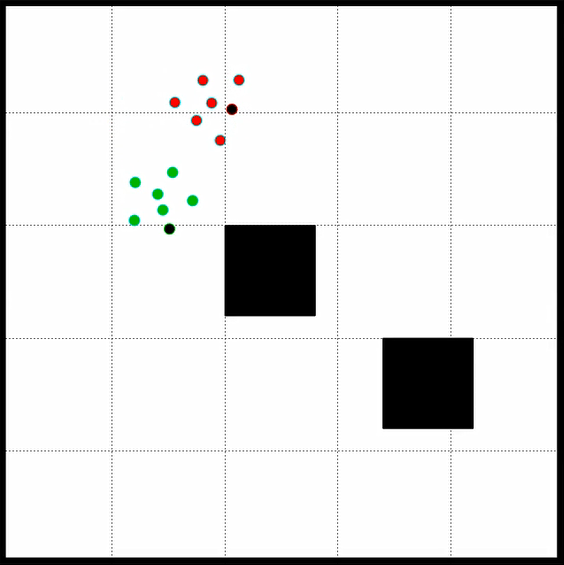}}
	\hspace{0.01\linewidth}
	\subfigure[]{
		\includegraphics[width=0.2\linewidth]{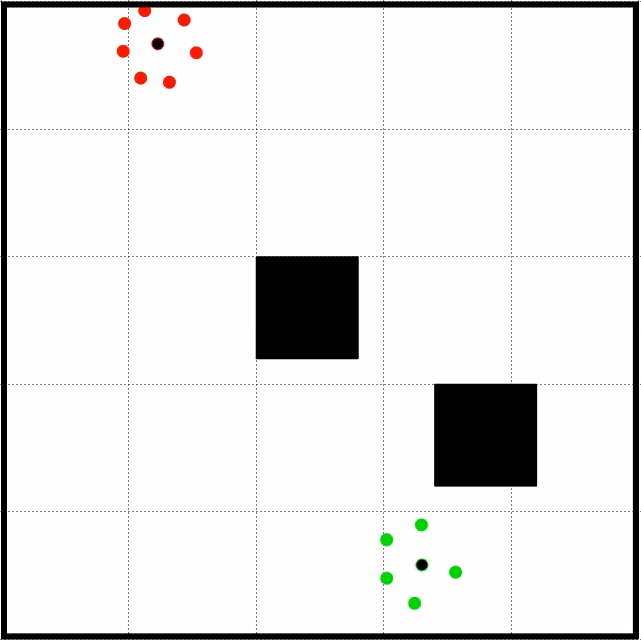}}
	\subfigure[]{
		\includegraphics[width=0.2\linewidth]{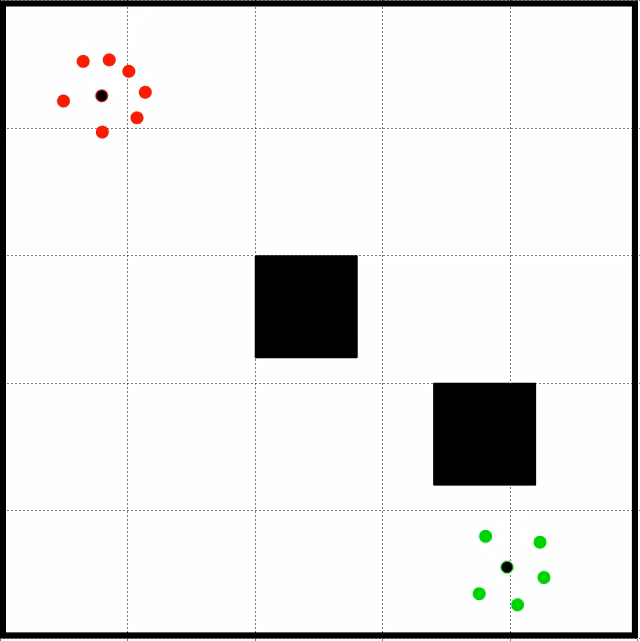}}
	\caption{Process of agents entrap two targets with GRN. (a) t=0s (b) t=14s (c) t=240s (d) t=266s.}
\end{figure*}\label{fig2}

\begin{figure*}[htbp]
	\centering
	\subfigure[]{
		\includegraphics[width=0.2\linewidth]{initial}}
	\hspace{0.01\linewidth}
	\subfigure[]{
		\includegraphics[width=0.2\linewidth]{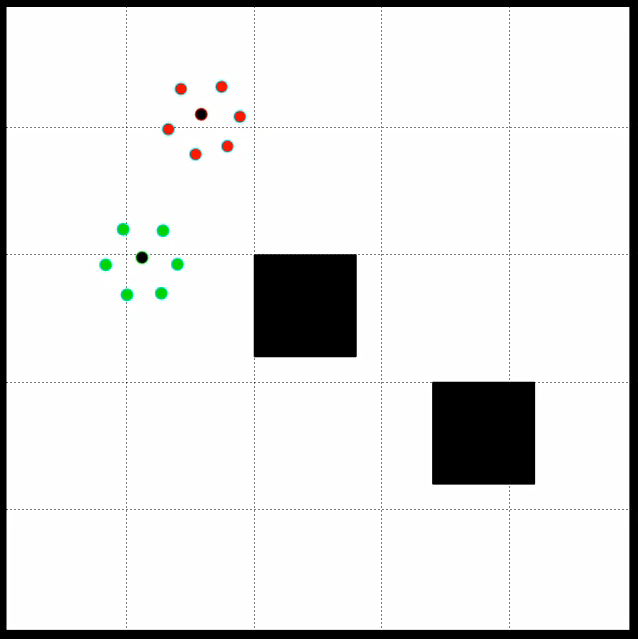}}
	\hspace{0.01\linewidth}
	\subfigure[]{
		\includegraphics[width=0.2\linewidth]{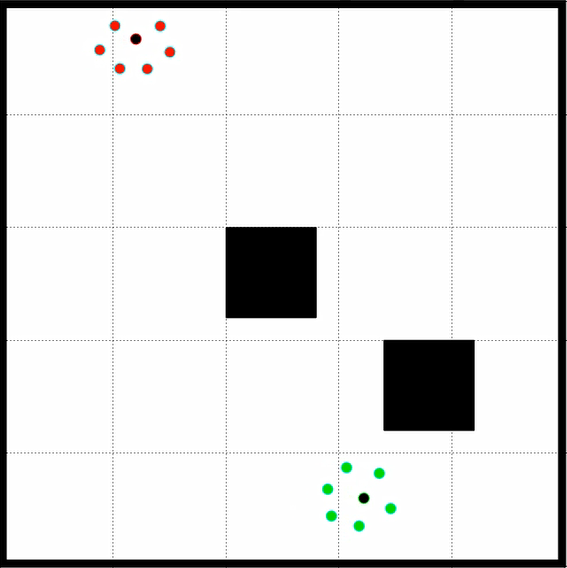}}
	\subfigure[]{
		\includegraphics[width=0.2\linewidth]{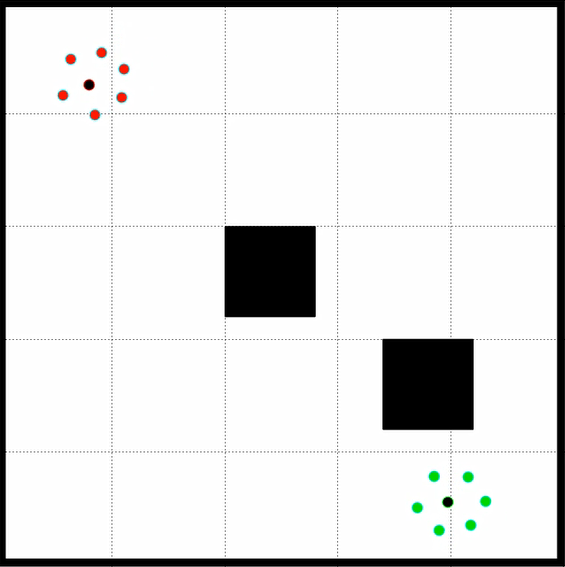}}	
	\caption{Process of agents entrap two targets with our method. (a) t=0s (b) t=14s (c) t=240s (d) t=266s.}
\end{figure*}	\label{fig3}

By comparing the pictures, both methods developed a formation that surrounded two targets to a certain extent. However, under the AGENT method, the number of agents was more uniform for each target in the arena, and the distribution of the agents' positions was more uniform. The AGENT method was more capable of dealing with multiple targets than the GRN method in entrapping multiple targets.

\subsection{Scalability Analysis}
To digitize the entrapping effect, we designed statistical indicators for the experiments.
This study calculates the corresponding entrapping indicators in experiments to compare the entrapping effects of the two methods. 
First, we expect that the position distribution of the agent around the target should be as even as possible.
In other words, the agents should not be too crowded in one direction of the target but should disperse as evenly as possible around the target. We define the occupancy rate of the encircling circle as the uniformity of the agent position around the target to evaluate the effect of entrapping.
Then, in the mission of entrapping multiple targets in a swarm, we expect agents to be nearly evenly distributed when the importance of the goal is the same. In this manner, multiple targets will be entrapped as perfectly as possible. 
In addition, the agents should entrap the target as soon as possible and with less distance, we calculate the response time and distance covered by the agents in entrapping targets. In addition, considering the agent flock to perform tasks, we measured the minimum distance between agents in the process of entrapping to show the safety of the two methods.
Lastly, in practical applications, the motion of the swarm should be as stable as possible when performing tasks, rather than suddenly changing directions. To meet the requirements of practical applications, we calculated the continuous velocity correlation of the agents in the entrapping process. 

According to the above principles, six statistical indicators for evaluating the effectiveness were designed as follows:

\begin{itemize}	
	\item The number of agents near each target.
	\item The uniformity of the agent position around the target.
	\item Time for agents to find one target and all targets.
	\item The average distance for agents entrapping all targets at firstly time.
	\item Agents' minimum distance of entrapping process.
	\item Velocity correlation of each agent's movement during entrapping.
\end{itemize}

In the arena, twelve agents entrapped two targets. The velocity of the targets was $ 1.8 m/step $. In our method, the velocity of the agents changed with time. The maximum velocity was $ 4 m/step $. In GRN, the velocity of the agents was $ 3.6 m/step $. 
In such a scenario, we compare the AGENT and GRN methods using the above indicators. 
The statistical results are shown in Fig.4 $\sim$ Fig.10. 

If six agents are assigned to each target, it is an ideal state for the index of the uniformity of the number of agents entrapping the target. However, owing to factors such as the real-time position of the agent from the target, the distribution may not be ideal. The two targets wandered on the map with steps conforming to the L\'{e}vy distribution. If the flexibility of the agents is insufficient, they may not be able to form a timely and even encirclement based on the flexible movement of the target with time. That is a challenge for agents. In the case of the same moving trajectory of targets, we carried out index statistics on the implementation of the entrapping mission. Firstly, we compare the number of agents allocated to each target by the AGENT and GRN methods. From Fig.4, the AGENT method we propose has better decision-making ability when entrapping multiple targets, that is, grouping more even in numbers.

\begin{figure}[htbp]
	
	\centering
	\includegraphics[width=9cm]{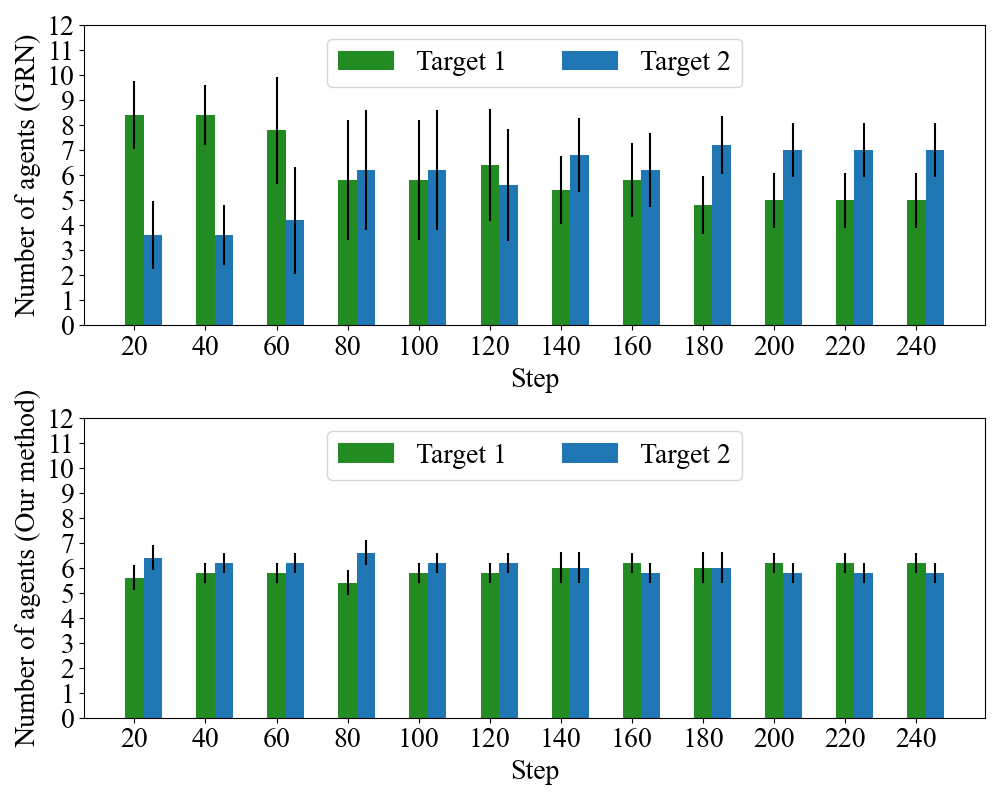}
	\caption{Number statistics of agents entrapping two targets. This is the statistical result of six simulation experiments (240 steps in every simulation experiment) with twelve agents and two targets in the same arena. Ideally, six agents should be assigned to each target. The figure shows the number of agents around the two targets at different sampling moments with the GRN and our AGENT method.}	
\end{figure}\label{fig4}

We the calculated the distribution of agents within a certain range in six directions around the target as the index of uniformity of the agent position around the target. In other words, an imaginary circle with the location of the target as the origin and radius of the distance that we recognize (such as 32 m) as the radius. We divide this circle evenly into six sectors. To test whether the agent entrapped the target, we counted the number of agents appearing in these sectors in the same experimental scene. From Fig.5, in the AGENT method, the agents are more evenly distributed within the encirclements they generate. This shows that the AGENT method is superior to the GRN method.

\begin{figure}[htbp]
	
	\centering
	\includegraphics[width=9cm]{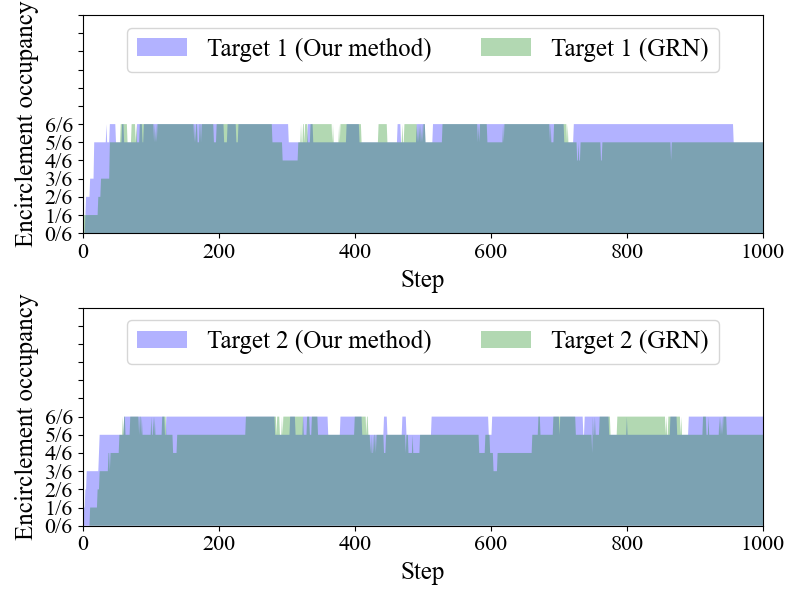}
	\caption{Statistics of the distribution of agents around the target. Twelve robots and two targets moved 1,000 steps in the arena using the GRN and our method. We divided a certain size(radius=32 m) circle of each target into six uniform fan-shaped areas, and counted the distribution of agents in the fan-shaped area with a sampling interval of one step.}
\end{figure}\label{fig5}

In this study, we assume that the encirclement occupancy of the target is $6/6$, indicating that the target is successfully encircled. In a real entrap task, agents need to achieve a uniform entrapping effect as quickly as possible. Therefore, this study counts the time that the agent first entered the all fan-shaped areas around one target and all targets in Fig.6. In addition, this study calculates the average distance of agents’ first entrance in all fan-shaped areas around all targets, as shown in Fig.7. It can be regarded as the average distance for agents entrapping all targets for the first time.

\begin{figure}[htbp]
	
	\centering
	\includegraphics[width=8cm]{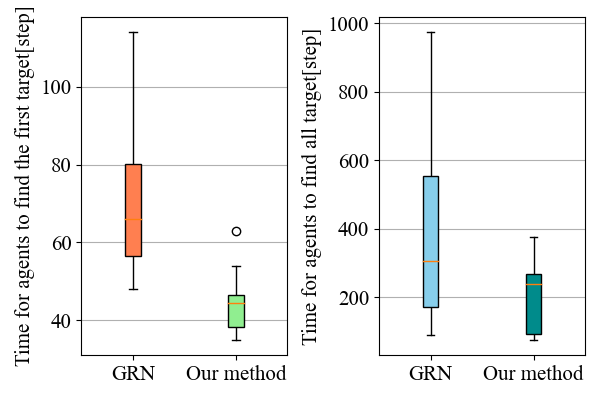}
	\caption{Time for agents to entrap one target and all targets for the first time. Twelve robots entrap two targets in the 250 m*250 m arena, using GRN and our method. The figure shows the time of agents occupying all fan-shaped areas around one target and all targets.}
\end{figure}\label{fig6}

\begin{figure}[htbp]
	
	\centering
	\includegraphics[width=8cm]{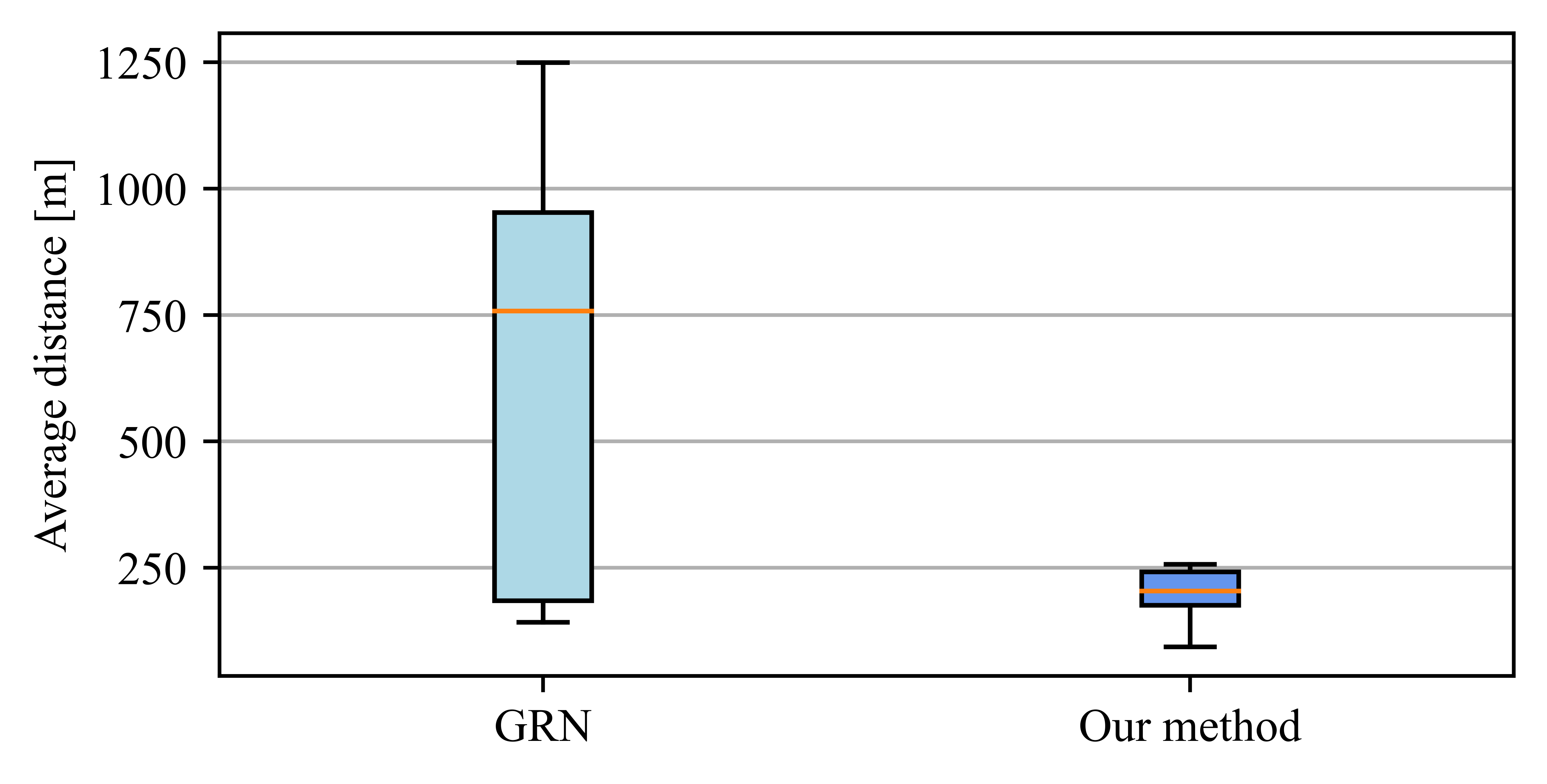}
	\caption{Average distance for agents to entrap all targets for the first time. Twelve robots entrap two targets in the 250 m*250 m arena, using GRN and our method. The figure shows the distance of agents occupying all fan-shaped areas around all targets.}
\end{figure}\label{fig7}

With the goal of providing safety of practical applications of agents flocking to performing entrapping tasks, this study calculates some evaluation indicators related to the motion of flocking. First, we do not expect the distance between agents to become too small. It would be dangerous while flocking in the real world. The average minimum distance was calculated as shown in Fig.8. The agents should make reasonable decisions to minimize sudden changes in the velocity direction. This increases the security of real-machine applications. The statistics are shown in Fig.9 and Fig.10, respectively. The formula for calculating the velocity correlation is as follows:

\begin{equation}	
	\phi^{\mathrm{corr}}= \frac{{v}_{i} \cdot {v}_{i-1}}{\left|{v}_{i}\right|\left|{v}_{i-1}\right|}
\end{equation}
where ${v}_{i}$ represents the velocity of the agent at time(step) $i$, and ${v}_{i-1}$ represents the velocity of the agent at the previous time(step) of ${v}_{i}$. As shown in Fig.9 and Fig.10, the angle between the current velocity of the agent and the velocity at the previous moment, the closer $\phi^{\mathrm{corr}}$ is to $1$, and the more stable the velocity direction of the agent is. Conversely, the closer $\phi^{\mathrm{corr}}$ is to $-1$, the more drastic the velocity direction of the agent changes. We can clearly see that the agent using our AGENT method could obtain a more stable velocity direction than the GRN method.
\begin{figure}[hpbp]
	
	\centering
	\includegraphics[width=8cm]{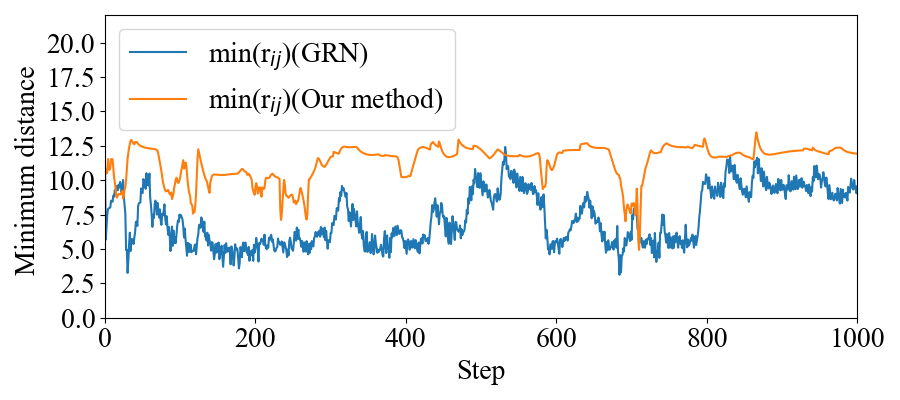}
	\caption{Agents' minimum distance of entrapping process. Twelve robots entrap two targets in the 250m*250m arena, with GRN and our AGENT method. The figure shows the minimum distance of the agents during the 1000 steps.}
\end{figure}\label{fig8}
\begin{figure*}[htbp]
	\centering
	\includegraphics[width=14cm]{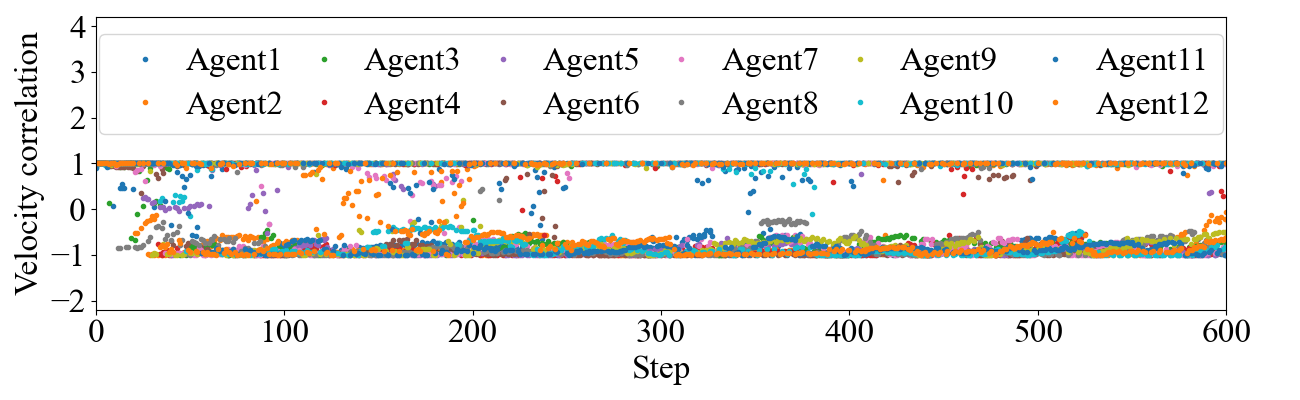}
	\caption{Velocity correlation of agents in GRN method at continuous time (step)}
\end{figure*}\label{fig9}
\begin{figure*}[htbp]
	\centering
	\includegraphics[width=14cm]{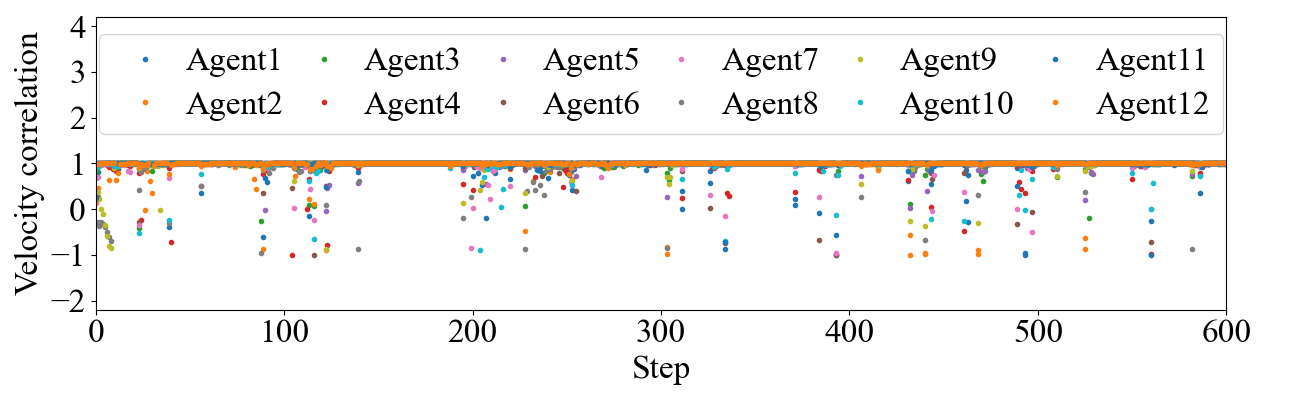}
	\caption{Velocity correlation of agents in AGENT method at continuous time(step)}
\end{figure*}\label{fig10}

When the agents use two methods to perform the entrapping task in the same scene, the AGENT method can ensure that the safe distance between agents is stable within an appropriate range, which is better than the GRN method as  the data in Fig.8 shows. The AGENT method with such a performance greatly improves the security during the flocking behavior. Furthermore, it can be seen from Fig.9 that the agents may have some repeated jumps in the GRN (continuously changing the velocity direction by a large margin). This phenomenon is unfavorable for practical applications. It wastes a lot of movement resources due to decision errors and this movement is also dangerous for robots, especially in swarms. In our method, this sudden change in velocity direction is reduced (because the target will suddenly change direction, this behavior of agents should not be completely eliminated), as shown in Fig.10. 

The above six indicators prove the stability and superiority of the AGENT method from different directions; in general, the AGENT method developed in this study allows the agent system to achieve a good group entrapping effect. 

\section{Real-world experiments}\label{sec5}

To perform real-world experiments on the AGENT method, we chose the E-puck2 robots to perform the entrap task. In the arena with random obstacles and targets, ten E-puck2 robots used the AGENT method to entrap two targets. The target moved in the arena with the L\'{e}vy flight algorithm as its step generation mechanism. The robots in the swarm communicated with each other via WiFi and obtained global information from the motion capture device above the arena, including the position information of other robots, obstacles, and the boundary of the arena. The information was used for the robot to make decisions (regarding the target to entrap) and to make movement speed adjustments (how to achieve an entrap effect). If the robots entered the area of the quintile circle within a certain radius of the target at the same time, the target was successfully rounded up. We counted the time (Table \ref{tab1}) taken by the E-puck2 robots to entrap the two targets in six experiments. From the table, we can conclude that E-puck2 robots often do not need too much time to complete the task using our method. We selected two representative experiments, as shown in Fig.11 and Fig.12(see our real-world experiment video link in the appendix).

\begin{table}[htbp]
	\caption{Statistics of the time taken by robots entrap two targets in six real-world experiments\label{tab1}}
	\centering
	\begin{tabular}{|cccccccccc|ccc|}
		\hline
		Experiments  & 1st  & 2nd & 3rd & 4th & 5th & 6th\\
		\hline
		Entrap one target   & 35s   & 47s  & 30s & 67s & 34s & 57s   \\
		\hline
		Entrap all targets    & 64s   & 52s  &  46s  &   72s  &  59s  &  67s  \\
		\hline
	\end{tabular}
\end{table}

\begin{figure*}[htbp]
	\centering
	\subfigure[]{
		\includegraphics[width=0.42\linewidth]{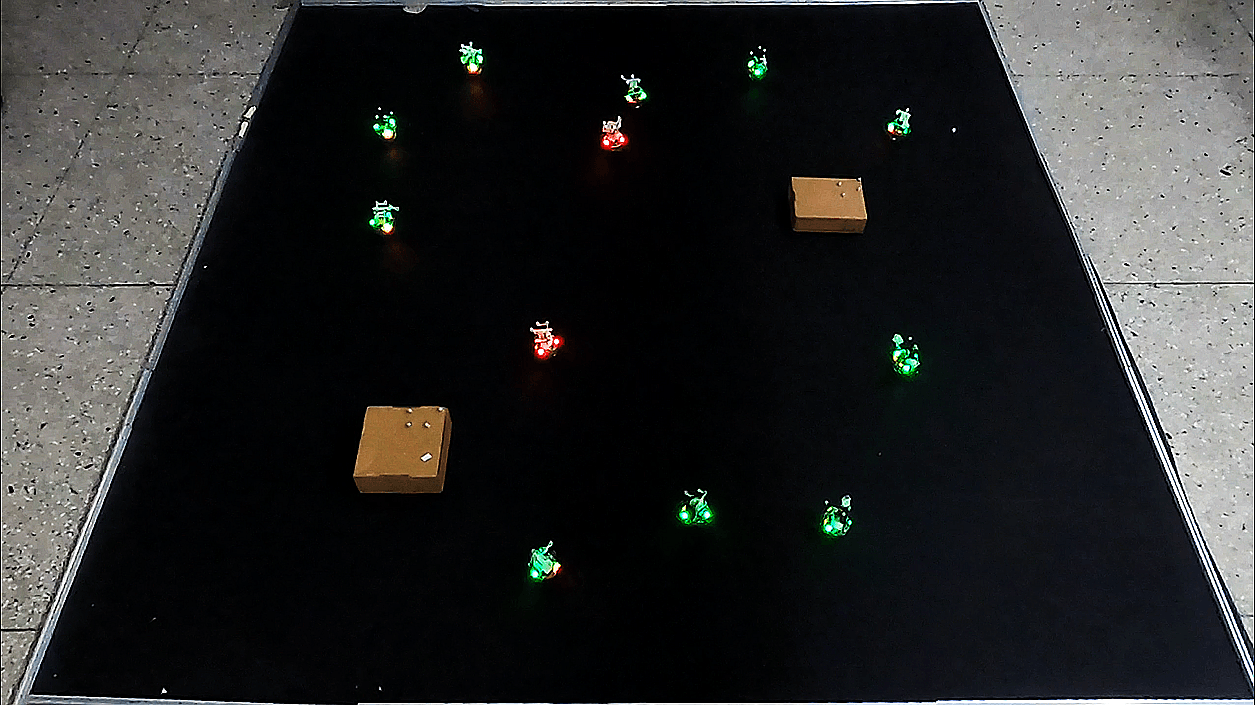}}
	\hspace{0.0005\linewidth}
	\subfigure[]{
		\includegraphics[width=0.42\linewidth]{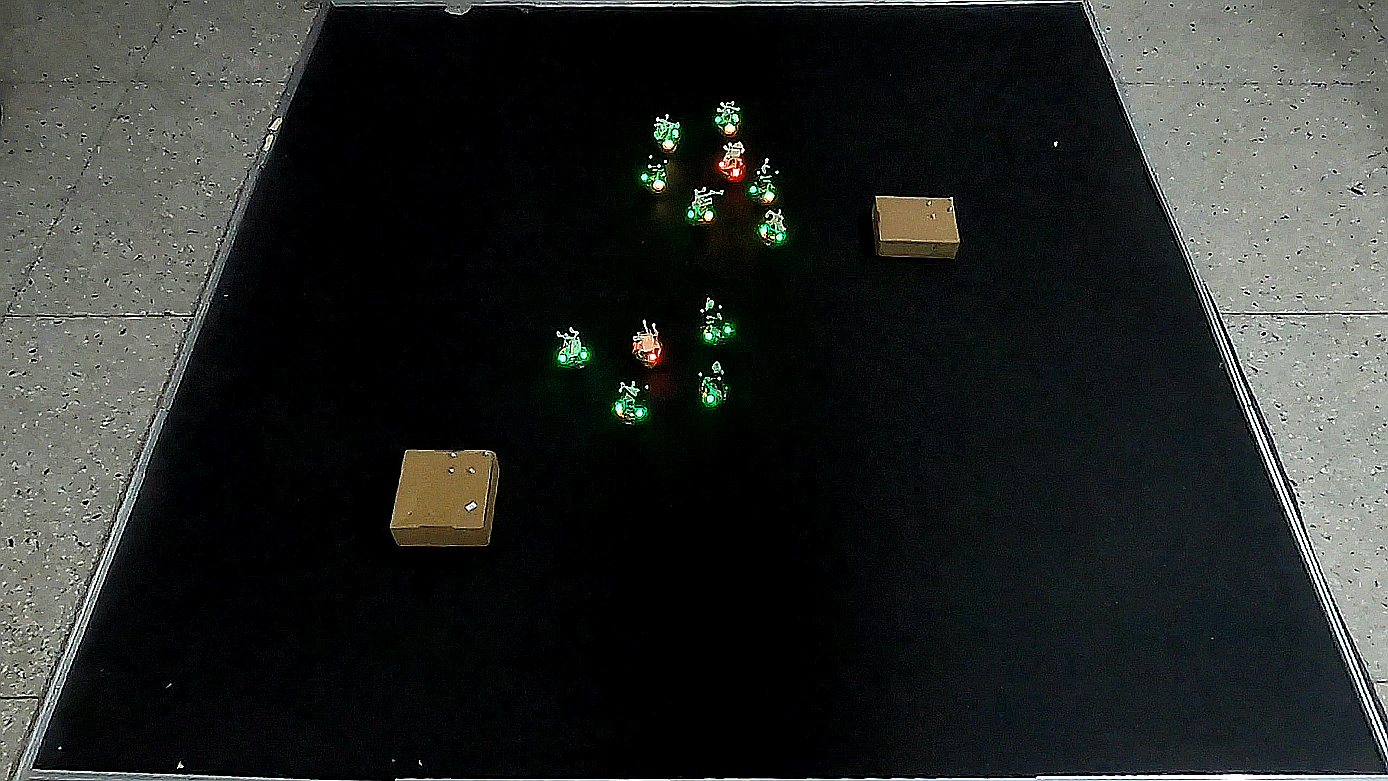}}
	\hspace{0.0005\linewidth}
	
	\subfigure[]{
		\includegraphics[width=0.42\linewidth]{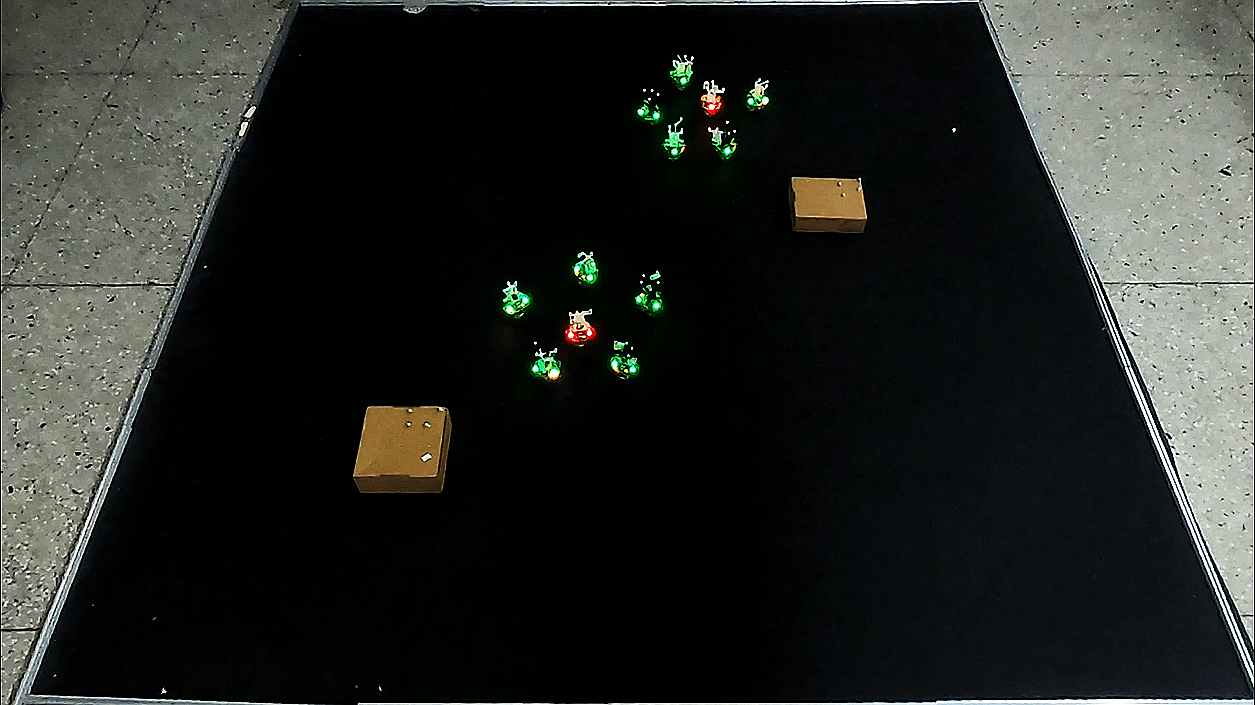}}
	\hspace{0.0005\linewidth}
	\subfigure[]{
		\includegraphics[width=0.42\linewidth]{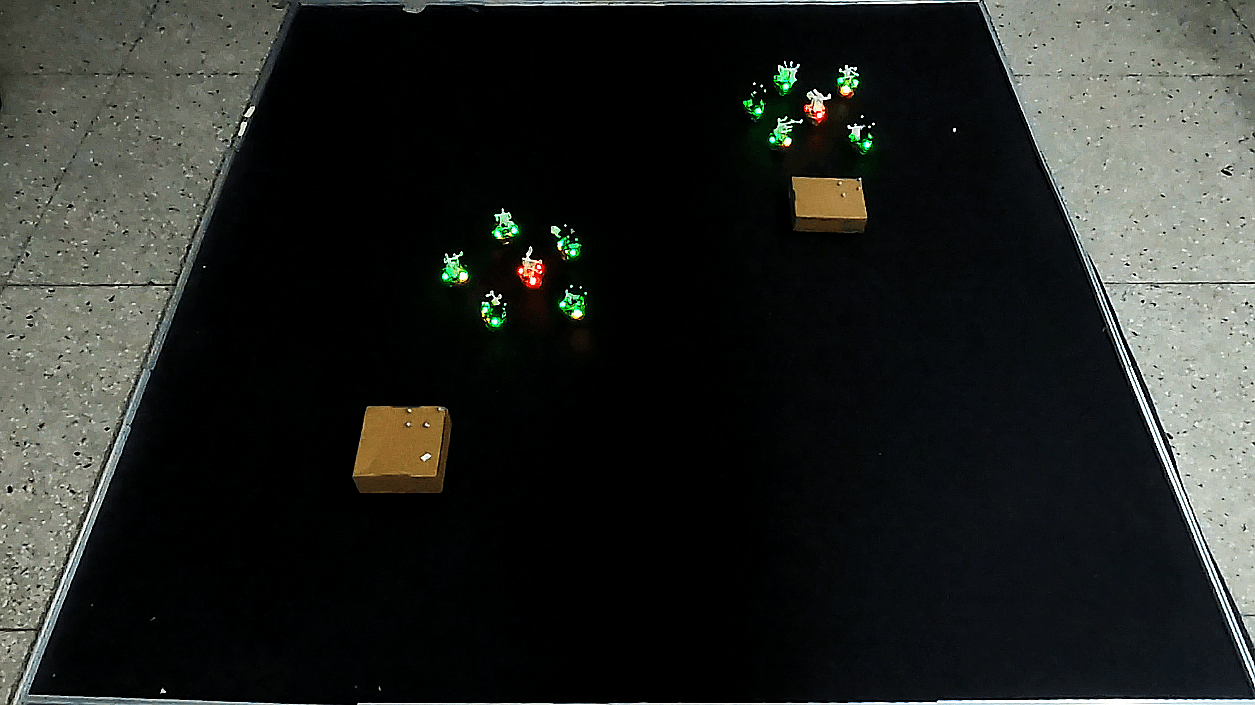}}
	\caption{Real-world experiments of agents entrapping two targets(scene1). Ten E-puck2 robots entrap two targets in a 3 m*3 m arena. (a) t=0 s (b) t=28  s (c) t=60s (d) t=132 s.}
\end{figure*}\label{fig11}
\begin{figure*}[htbp]
	\centering
	\subfigure[]{
		\includegraphics[width=0.42\linewidth]{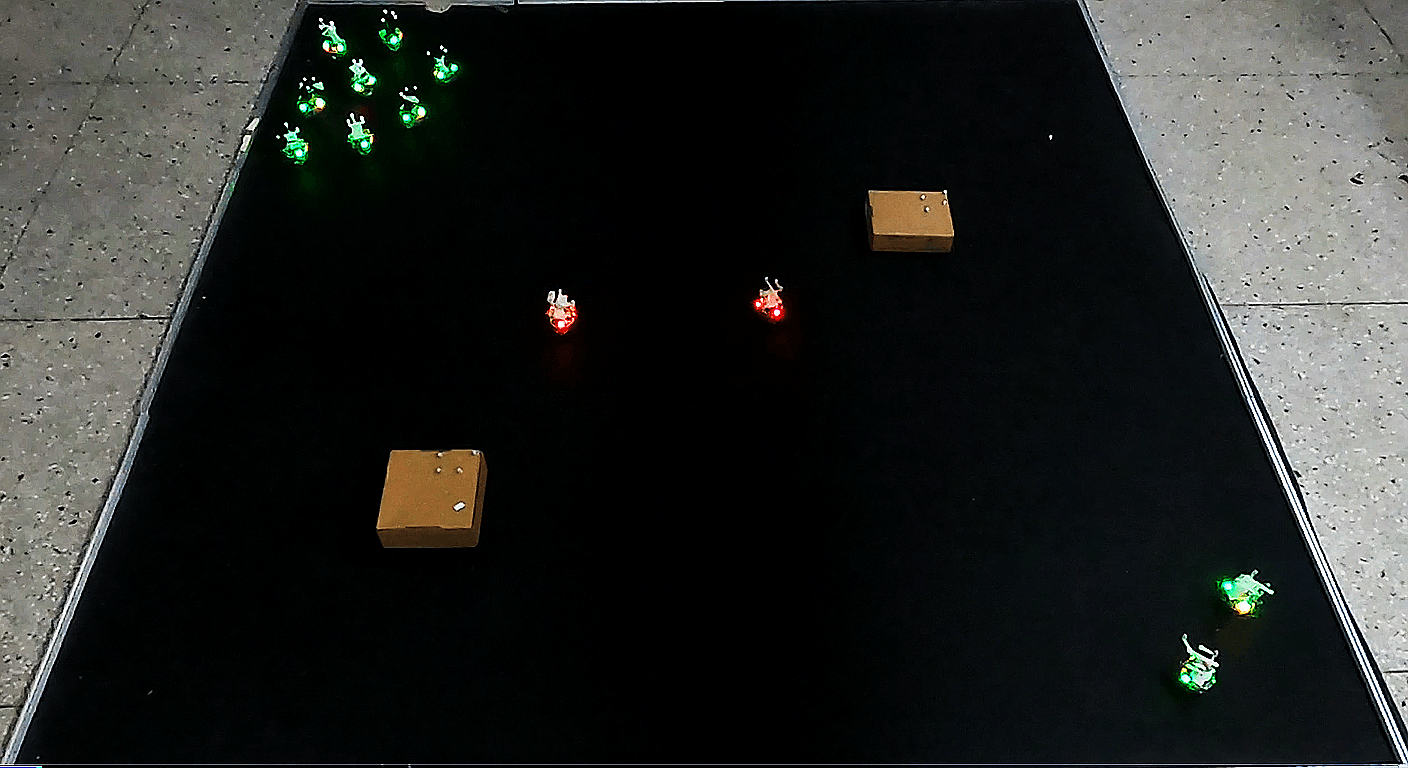}}
	\hspace{0.0005\linewidth}
	\subfigure[]{
		\includegraphics[width=0.42\linewidth]{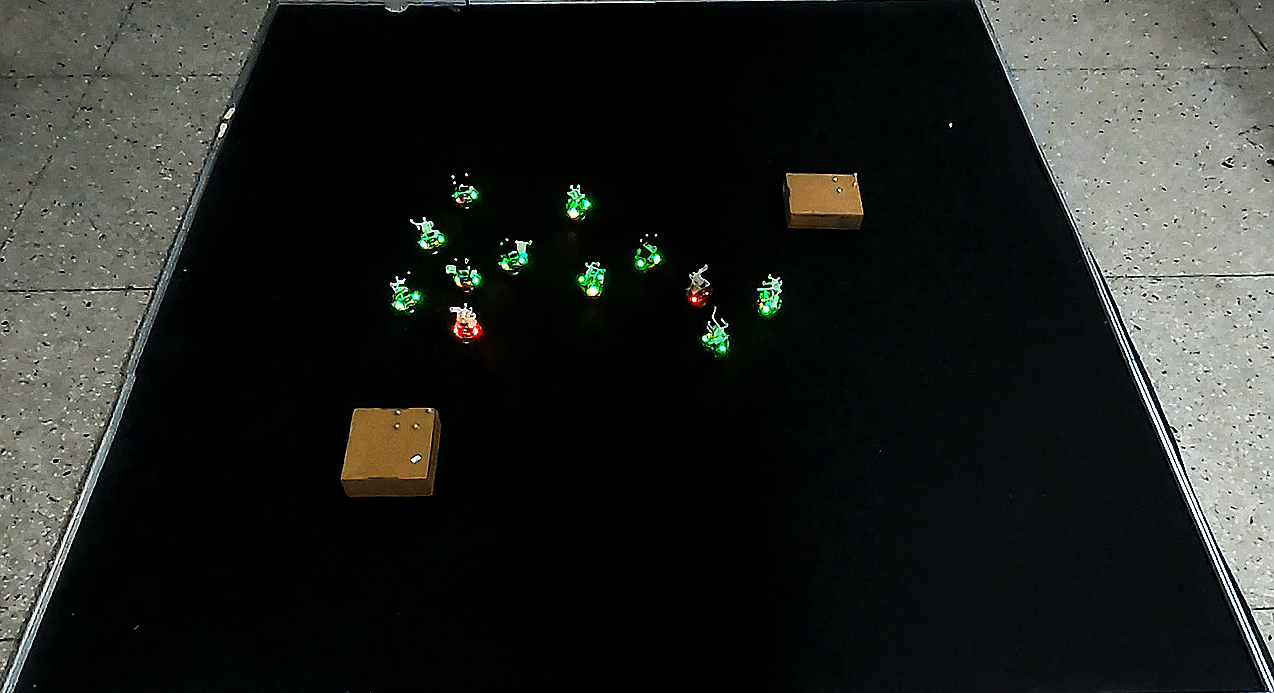}}
	\hspace{0.0005\linewidth}
	
	\subfigure[]{
		\includegraphics[width=0.42\linewidth]{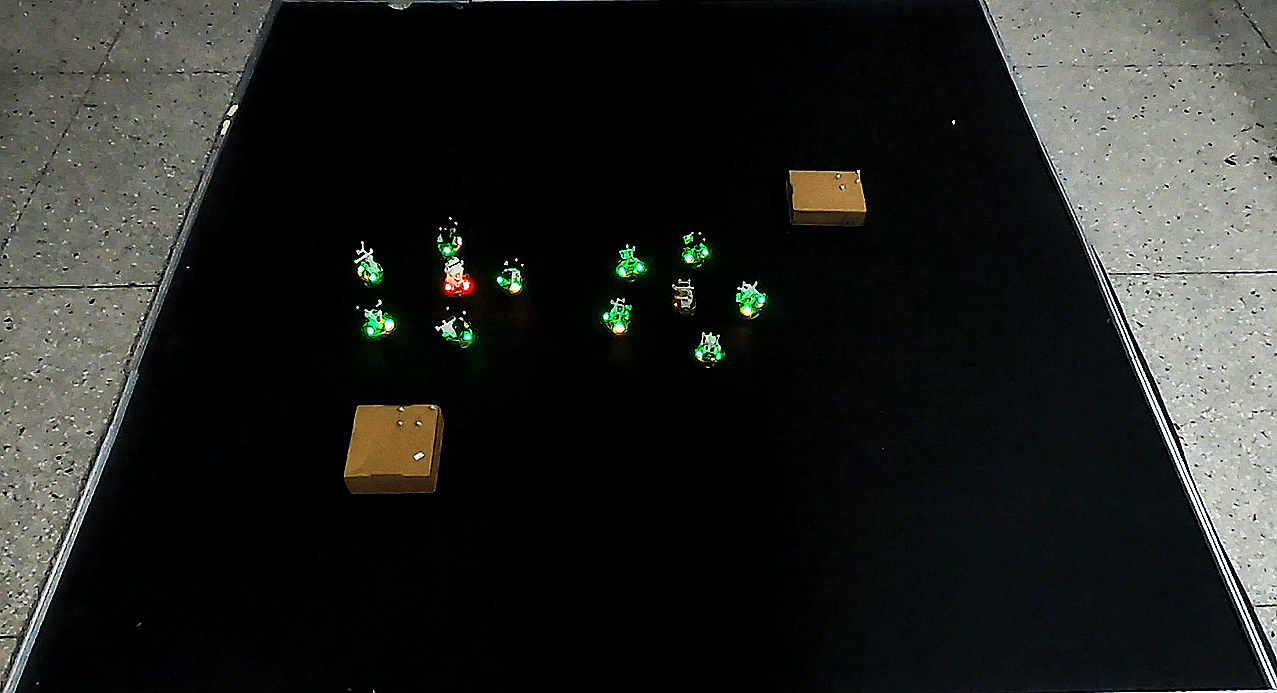}}	
	\hspace{0.0005\linewidth}
	\subfigure[]{
		\includegraphics[width=0.42\linewidth]{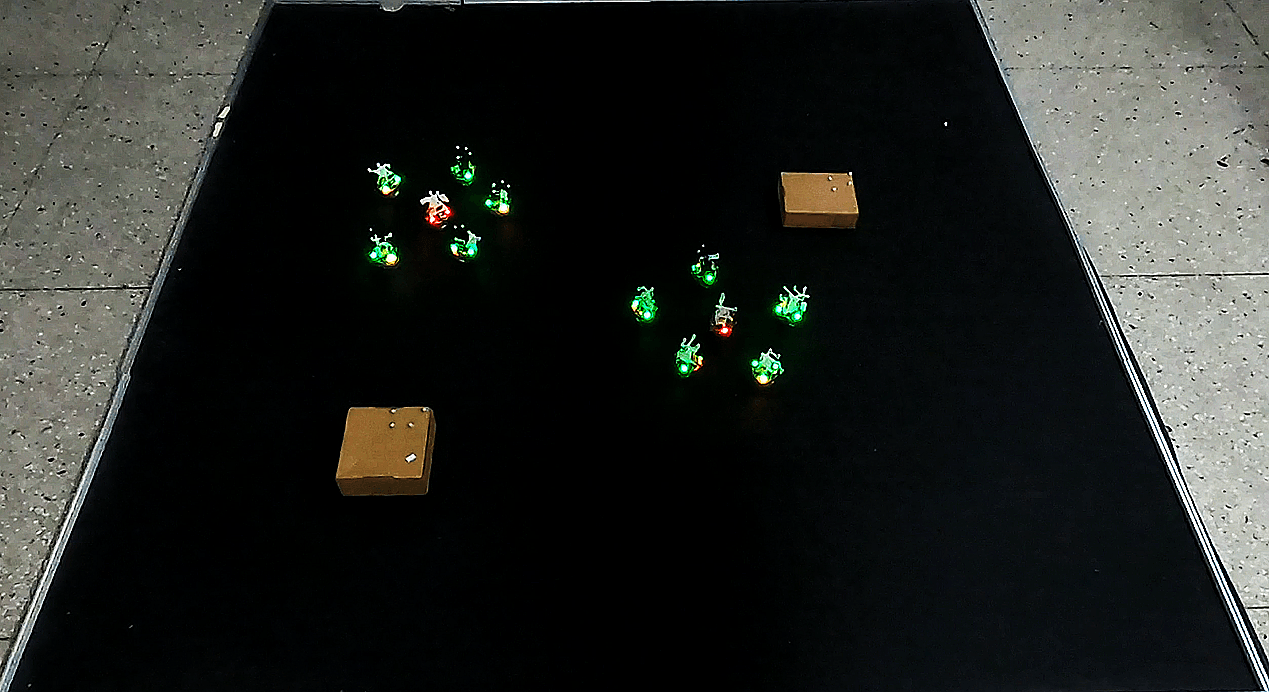}
	}
	\caption{Real-world experiments of agents entrapping two targets(scene2). Ten E-puck2 robots entrap two targets in a 3m*3m arena. (a) t=0s (b) t=35s (c) t=67s (d) t=139s. }\label{fig12}
\end{figure*}
As shown in Fig.11 and Fig.12, in the case of the scattered distribution of the robots' positions, E-puck2 robots can adaptively group and entrap the target evenly according to the environmental conditions, flexibly adjust the formation to adapt to the environment, and do not collide with neighbors and obstacles. Even if the target suddenly changes direction, the agent adaptively adjusts its speed to meet the target to form a tight circle. 
The real-world experiment verifies the effectiveness of AGENT method. 
\section{CONCLUSIONS}\label{sec6}
This study proposes an adaptive grouping method to entrap multiple targets with distributed systems.
Using our method, the agent can make decisions in real-time based on environmental information, resulting in the effect of an even grouping around the target. The agents can flexibly respond to a sudden change in the direction of the target and always adapt to the change in movement speed to maintain the entrapping formation of the target. From a security point of view, the AGENT method considers the kinematic constraints of the agents. There were no collisions between the agents and obstacles in the flocking system. Furthermore, the flocking movement looks as smooth and natural as possible, although environmental factors are complex and changeable.
Simulation experiments and index analyses prove that the AGENT method improves the grouping entrapping effect of the swarm system compared to the GRN method. Finally, real-world experiments with E-puck2 robots are presented to demonstrate the effect of the AGENT method.

\appendix
Simulation experiment with AGENT entrapping system: \url{https://www.bilibili.com/video/BV1sa411r72q?spm_id_from=333.999.0.0}

Real-world experiment with AGENT entrapping system: \url{https://www.bilibili.com/video/BV1TZ4y197E3?spm_id_from=333.999.0.0}



\clearafterbiography\relax

\end{document}